\documentclass[pra,twocolumn,showpacs]{revtex4}
\usepackage{amsmath}
\usepackage{graphicx}
\usepackage{amssymb} 
 
\begin{document} 

\title{Entangled Quantum States Generated by Shor's Factoring Algorithm}
\date{\today} 
 
\author{Yishai Shimoni, Daniel Shapira and Ofer Biham} 
\affiliation{Racah Institute of Physics, 
The Hebrew University, Jerusalem 91904, Israel} 
 
\begin{abstract} 

The intermediate quantum states
of multiple qubits,
generated during the operation of
Shor's factoring algorithm are analyzed. 
Their entanglement is evaluated using the Groverian measure.
It is found that the entanglement is generated during the
pre-processing stage of the algorithm and remains nearly constant
during the quantum Fourier transform stage.
The entanglement is found to be correlated with the speedup 
achieved by the quantum algorithm compared to classical 
algorithms.

\end{abstract} 
 
 
\maketitle 
 
\section{Introduction} 
\label{sec:introduction}

The potential speedup of quantum algorithms is demonstrated
by Shor's factoring algorithm, which is exponentially
faster than any known classical algorithm
\cite{Shor1994}.
Several other quantum algorithms, which 
are more efficient than their classical counterparts
were introduced 
~\cite{Deutsch1992,Grover1996,Grover1997,Jozsa1997}. 
Factorization is of special interest 
due to its role in
current methods of cryptography.
Although the origin of 
the speed-up offered by quantum algorithms is not fully understood, 
there are indications that quantum entanglement 
plays a crucial role
\cite{Jozsa2003,Vidal2003}. 
In particular,
it was shown that quantum algorithms that do not create entanglement 
can be simulated efficiently on a classical computer
\cite{Aharonov1996}.
It is therefore of interest to quantify the entanglement 
produced by quantum algorithms and examine its correlation
with their efficiency.  
This requires to develop entanglement measures for the quantum states
of multiple qubits that appear in quantum algorithms.
Recently, the Groverian measure of entanglement was introduced and
used for the evaluation of entanglement in certain pure quantum states
of multiple qubits
\cite{Biham2002}.
Using computer simulations of the
evolution of quantum states during the operation of 
a quantum algorithm, 
one can obtain the time evolution of the entanglement.
Such analysis was performed 
for Grover's search algorithm
with various initial states and different choices of the marked states
\cite{Shimoni2004}.   
It was shown that Grover's iterations generate highly entangled
states in intermediate stages of the quantum search process,
even if the initial state and the target state are product states. 

In this paper we analyze the quantum states that are
created during the operation of Shor's factoring algorithm.
The entanglement in these states is evaluated using the
Groverian measure.
It is found that the entanglement is generated during the
pre-processing stage.
When the quantum Fourier transform (QFT) is applied to the
resulting states, their entanglement remains unchanged.
This feature is unique to periodic quantum 
states, such as those that result from the
pre-processing stage of Shor's algorithm.
When other states, such as product states or random states
are fed into the QFT, their entanglement does change.
Another interesting feature is that
the entanglement is found to be correlated with the speedup 
achieved by the quantum factoring algorithm compared to classical 
algorithms.
This means that the cases where no entanglement is created
are those in which classical factoring is efficient. 

The paper is organized as follows.
In Sec. 
\ref{sec:algorithm} 
we briefly review Shor's factoring algorithm,
the QFT algorithm, and the quantum circuit used to perform it.
In Sec. 
\ref{sec:groverian} 
we describe the Groverian entanglement measure
and the numerical method in which it is calculated.
In Sec. 
\ref{sec:ent} 
we use the Groverian measure to evaluate
the entanglement created by Shor's algorithm.
The results are discussed in Sec.
\ref{sec:discussion}
and summarized in Sec.
\ref{sec:summary}.

\section{Shor's Factoring Algorithm}
\label{sec:algorithm}


Shor's algorithm factorizes a given non-prime integer $N$,
namely, it finds integers $p_1$ and $p_2$, 
such that their product 
$p_1 p_2  = N$.
The algorithm consists of three parts:
(a) Pre-processing stage, in which 
the quantum register is prepared using classical algorithms
and quantum parallelism;
(b) Quantum Fourier transform, which is applied on the 
output state of the previous stage;
(c) Measurement of the register and post-processing
using classical algorithms.

\subsection{Pre-processing}

Given an integer $N$ to be factorized, 
choose any integer $y<N$,
and find the
integer $q=2^L$ that satisfies 

\begin{equation}
N^2 < q \leq 2 N^2.
\label{eq:<q<}
\end{equation}

\noindent
Prepare a register of
$L$ qubits (later referred to as the main register)
in the equal superposition state

\begin{equation}
| \eta \rangle = \frac{1}{\sqrt{q}} \sum_{a=0}^{q-1} | a \rangle.
\end{equation}

\noindent
Next, use quantum operations to calculate
$y^a \ {\rm mod}\ N$ 
for all the indices, 
$a=0,\dots,q-1$,
of the basis states above,  
and store the results in an auxiliary register, 
giving rise to the joint state

\begin{equation}
\frac{1}{\sqrt{q}} \sum_{a=0}^{q-1} | a \rangle |y^a \ {\rm mod}\ N \rangle.
\end{equation}

\noindent
This essentially completes the pre-processing stage.
However, in order to present the next stage of the algorithm more clearly,
it is helpful to measure the auxiliary register in the computational
basis.
Suppose that the result of the measurement is 
a state $| z \rangle$, where
$z = y^l \ ({\rm mod}\ N)$ 
and $l$ is the smallest positive integer that gives the value $z$.
The order of $y$ modulus $N$ is defined as an integer $r$ that
satisfies $y^r = 1 \ ({\rm mod}\ N)$.
The equality

\begin{equation}
y^{jr+l} = y^l \ \ ({\rm mod}\ N)
\label{eq:repitition}
\end{equation}

\noindent
is thus satisfied
for any integer $j$.
From 
Eq.~(\ref{eq:repitition}) 
it follows that
the measurement will select from the 
main register all values of
$a=l,l+r,l+2r,\ldots,l+Ar$,
where $A$ is the largest integer which is smaller than $(q-1)/r$.
The state of the register after the measurement is therefore

\begin{equation}
| \phi_l \rangle = \frac{1}{\sqrt{A+1}} \sum_{j=0}^A |jr+l\rangle.
\label{eq:phi_l}
\end{equation}

\subsection{Quantum Fourier Transform}

\begin{figure}
\includegraphics[width=8.5cm]{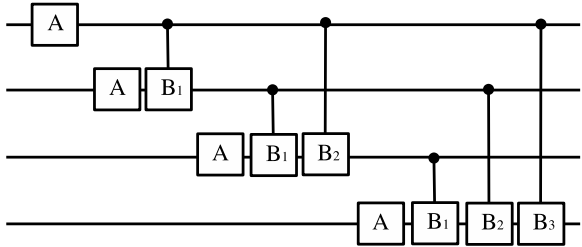}
\caption{The circuit of the quantum Fourier transform
(QFT) performed on a $4$-qubit register. 
The operator $A$ is the Hadamard gate.
The operators 
$B_{1}$,
$B_{2}$
and
$B_{3}$
are the controlled-phase gates
$B_{k,m}$,
where $m-k=1$, 2 and 3, respectively.
}
\label{fig:1}
\end{figure}

The quantum Fourier transform 
is given by

\begin{equation}
\sum_{a=0}^{q-1} f(a) |a\rangle \mapsto \sum_{c=0}^{q-1} \tilde{f}(c)|c\rangle,
\label{eq:QFT1}
\end{equation}

\noindent
where

\begin{equation}
\tilde{f}(c)=\frac{1}{\sqrt{q}}
\sum_{a=0}^{q-1} \exp\left(\frac{2\pi iac}{q}\right) f(a).
\end{equation}

\noindent
The quantum circuit of the QFT is shown in Fig. 
\ref{fig:1}.
To obtain the transformation in Eq.
(\ref{eq:QFT1}),
the $L$ qubits of register 
$| a \rangle$
in the input (and throughout the quantum circuit) are
indexed 
by $k=1,\dots,L$,
from bottom to top.
The output of the circuit
is stored in register
$| c \rangle$,
whose qubits 
are indexed from top to bottom.
We define the operator 
$A_k$ to be the Hadamard gate
applied to qubit $k$,
and the operator 
$B_{k,m}$ 
(where $m>k$)
to be a controlled phase operator,
which applies a phase of 
$\theta_{k,m}=\pi/2^{m-k}$
only if both qubits $k$ and $m$ are $1$.
We also define

\begin{equation}
F_k = A_k B_{k,k+1} B_{k,k+2} \dots B_{k,L},
\label{eq:F_j}
\end{equation}

\noindent
for $k=1,\dots,L$,
where we follow the standard notation for quantum operators,
namely, those on the right hand side operate first.
With these definitions the 
sequence of quantum operations that perform the QFT
is given by

\begin{equation}
{\rm QFT} = F_1 F_2 \dots F_L.
\end{equation}

\noindent
The number of one-qubit and two-qubit gates required 
in the quantum circuit which performs QFT
is polynomial in the size of the register.

In the simple case in which
$r$ divides $q$ exactly,
namely
$A+1=q/r$, 
one obtains

\begin{equation}
{\rm QFT} |\phi_l\rangle=
\frac{1}{\sqrt{r}}\sum_{j=0}^{r-1} \exp\left(\frac{2\pi ilj}{r}\right)
\left|j\frac{q}{r}\right\rangle.
\label{eq:QFT}
\end{equation}

\noindent
where 
$|\phi_l \rangle$
is defined in Eq.
(\ref{eq:phi_l}).
The resulting state is a superposition of 
all basis states with indices which are products of $q/r$.
If $r$ is not a divisor $q$, namely, 
$q/r$ is not an integer,
Eq.~(\ref{eq:QFT})
should be modified such that 
the large amplitude states are 
those which correspond to integers
adjacent to
$j q/r$, $j=0,1,\dots,r-1$.
Our choice of $q$ in 
Eq.~(\ref{eq:<q<}) 
ensures that with high probability
the measurement will yield only states
whose indices are the nearest integers to
$j q/r$.

\subsection{Measurement and Post-Processing}

The third part of the algorithm starts with a 
measurement of the register.
It yields an integer approximation, $c$, of one of the 
values $j q/r$, $j=0,1,\dots,r-1$.
Thus, $cr$ is approximately an integer multiple of $q$.
Here, again, our choice of $q$ in 
Eq.~(\ref{eq:<q<}) 
ensures that
in most cases there exist another integer $c'$
which satisfies
$|rc-c'q|\leq r/2$. 
As a result

\begin{equation}
\left|\frac{c}{q}-\frac{c'}{r}\right|\leq\frac{1}{2q}.
\label{eq:approx}
\end{equation}

\noindent
Using a continued fraction expansion of 
$c/q$ it is possible to efficiently
find $c'$ and $r$. 
There is only one such approximation which satisfies 
Eq.~(\ref{eq:approx})
for $r<N$. 
Thus, the correct value of $r$ is obtained.
If $r$ is even we can define
$x=y^{r/2}$
which satisfies

\begin{equation}
x^2-1=(x-1)(x+1) = 0 \ ({\rm mod}\ N).
\label{eq:x^2-1}
\end{equation}

\noindent
From 
Eq.~(\ref{eq:x^2-1}) 
we obtain that 
$x+1 \ ({\rm mod}\ N)$ 
and 
$x-1 \ ({\rm mod}\ N)$
are candidates for having a common divisor with $N$.
Using Euclid's 
greatest common divisor (GCD)
algorithm,
this common divisor 
is found and the factoring process is completed.

\section{The Groverian Measure of Entanglement}
\label{sec:groverian}

\subsection{Formal Definition}

Consider a quantum algorithm,
given by the unitary operator  
$U$, 
applied to the equal superposition 
state 
$| \eta \rangle$.
For a certain class of quantum algorithms, 
the final, or target state

\begin{equation}
| t \rangle = U | \eta \rangle,
\label{eq:m=Ae}
\end{equation}

\noindent
is a computational basis state.
This state stores the correct result of the calculation,
which can be extracted by measurement.
Not all quantum algorithms can be expressed in this form,
because the final state,
before the measurement is done, 
may be a superposition state.
However, 
in the case of 
Grover's search algorithm with a single marked state,
this description applies
\cite{Biham2002}. 
Consider the case in which such algorithm, $U$, 
is applied to an arbitrary pure state 
$| \psi \rangle$. 
The probability of success
is defined as the probability that the measurement will 
still give the state
$| t \rangle$.
This probability is given by
$P_s  = |\langle t  | \psi \rangle |^2$.

The success probability can be used to evaluate the entanglement
of the state 
$| \psi \rangle$. 
To this end, 
before the algorithm $U$ is applied,  
one applies a local unitary operator, $U_k$,
on each qubit $k=1,2,\dots,L$. 
These operators are chosen such that the success probability of
the algorithm will be maximized.
The maximal success probability is

\begin{equation}
P_{\max}=\max_{U_1,\dots,U_L}
\left|\langle t|UU_1\otimes\dots\otimes U_L| \psi \rangle\right|^2.
\end{equation}

\noindent
Using 
Eq.~(\ref{eq:m=Ae}) 
the success probability
$P_{\max}$ 
can be expressed by

\begin{equation}
P_{\max}=\max_{U_1,\dots,U_L}
\left|\langle\eta |U_1\otimes\dots\otimes U_L| \psi\rangle\right|^2.
\end{equation}

\noindent
This can be re-written as

\begin{equation}
P_{\max}=\max_{|e_1\rangle,\dots,|e_L\rangle}
\left|\langle e_1 \otimes \dots \otimes e_L| \psi \rangle \right|^2,
\label{eq:Pmax}
\end{equation}

\noindent
where the
$|e_k\rangle$'s 
are single-qubit states.
Eq.~(\ref{eq:Pmax}) 
means that 
for a given initial state 
$| \psi \rangle$,
the maximal success probability 
of such algorithm, $U$, 
is equal to the maximal overlap of
$| \psi \rangle$
with any product state.

The Groverian measure of entanglement 
$G(\psi)$
is defined by

\begin{equation}
G(\psi) = \sqrt{1-P_{\max}}.
\end{equation}

\noindent
For the case of pure states, for which $G(\psi)$
is defined, it is closely related to an entanglement measure 
introduced in Refs.
\cite{Vedral1997,Vedral1997a,Vedral1998} and
was shown to be an entanglement monotone.
The latter measure is defined for both pure and mixed 
states. 
It can be interpreted as the distance 
between the given state and the nearest separable state
and expressed in terms of the fidelity of the
two states.
Based on these results, it was shown 
\cite{Biham2002}
that $G(\psi)$
satisfies:
(a) $G(\psi) \geq 0$, with equality only when $|\psi\rangle$
is a product state;
(b) $G(\psi)$ cannot be increased using local operations 
and classical communication
(LOCC). 
Therefore, $G(\psi)$ is an entanglement monotone
for pure states.
A related result was obtained in Ref.
\cite{Miyake2001},
where it was shown that the evolution of the quantum state  
during the iteration of Grover's
algorithm corresponds
to the shortest path in Hilbert space
using a suitable metric.

\subsection{Numerical Evaluation}

Consider a pure quantum state of $L$ qubits

\begin{equation}
|\psi\rangle=\sum_{j=0}^{2^L-1}a_j|j\rangle.
\end{equation}

\noindent
In order to find $G(\psi)$
we form a convenient representation of the tensor product states
used in Eq.
(\ref{eq:Pmax}).
The state of each qubit in the product state
is given by

\begin{equation}
|e_k\rangle = 
e^{i\delta_k}\left[\cos\theta_k|0\rangle+
e^{i\gamma_k}\sin\theta_k|1\rangle\right].
\label{eq:e_j}
\end{equation}

\noindent
Let us denote

\begin{equation}
b_j^{(k)}=\left\{
\begin{array}{ll}
\cos\theta_k & {\rm if} \, j_k=0 \\
e^{i\gamma_k}\sin\theta_k & {\rm if} \, j_k=1,
\end{array}
\right.
\end{equation}

\noindent
where 
$j_k$, $k=1,\dots,L$ 
is the $k$'th most significant bit in the binary 
representation of $j$.
The overlap 
between
$| \psi \rangle$
and the product state
$| e_1 \otimes \dots \otimes e_L \rangle$
is given by
$f(\psi,\theta_1,\dots,\theta_L,\gamma_1,\dots,\gamma_L) = 
\langle e_1 \otimes\dots\otimes e_L | \psi \rangle$.
It can then be written as

\begin{equation}
f(\psi,\theta_1,\dots,\theta_L,\gamma_1,\dots,\gamma_L)  
= \sum_{j=0}^{2^L-1} b_j^{(1)} b_j^{(2)}\dots b_j^{(L)} a_j.
\label{eq:foverlap}
\end{equation}

\noindent
The phases $\delta_k$ only introduce a 
global phase which can be ignored.
The Groverian entanglement measure 
for the state 
$| \psi \rangle$
is given by

\begin{equation}
P_{\rm max} = \max_{\theta_1,\dots,\theta_L,\gamma_1,\dots,\gamma_L}
\left|f(\psi,\theta_1,\dots,\theta_L,\gamma_1,\dots,\gamma_L) \right|^2,  
\end{equation}

\noindent
namely, the dimension of the parameter space in which the
maximization is obtained is $2L$.
However, the number of terms summed up in the 
calculation of $f$ increases exponentially with
the number of qubits.
Therefore, 
to make the calculation of $G(\psi)$ feasible one should 
minimize the number of evaluations
of $f$.
The commonly used 
steepest descent algorithm,
requires a large number of evaluations of $f$
and is thus computationally inefficient.
Here we accelerate the calculation by
performing the maximization analytically and separately 
for a single pair of  
$\theta_k$ and
$\gamma_k$. 
During each maximization step, all the other parameters are held fixed. 
In the maximization 
we have a function of
the form

\begin{equation}
f = c_k \cos \theta_k + d_k e^{i \gamma_k} \sin \theta_k,
\end{equation}

\noindent
where 
$a_k = |a_k| e^{i \alpha_k}$ 
and 
$b_j = |b_j| e^{i \beta_j}$  
depend on the other $2L-2$ parameters.
The maximization of $|f|^2$ vs. 
$\theta_k$ 
and
$\gamma_k$ 
leads to

\begin{equation}
|f|^2 \rightarrow |c_k|^2 + |d_k|^2
\end{equation}

\noindent
where

\begin{equation}
\cos \theta_k \rightarrow  \frac{|c_k|}{\sqrt{|c_k|^2 + |d_k|^2}}
\end{equation}

\noindent
and

\begin{eqnarray}
\gamma_k \rightarrow \alpha_k - \beta_k.
\end{eqnarray}

\noindent
Using this method, the number of evaluations of $f$
is significantly reduced.
To find the global maximum, $P_{\rm max}$ and then
$G(\psi)$
we perform several rounds of maximization over all the
$2L$ parameters.
Trying different initial conditions we find that the
convergence to the global maximum is fast and no other
local maxima are detected.

\section{Entanglement During Shor's Algorithm}
\label{sec:ent}

Shor's factoring algorithm includes
a pre-processing stage followed by QFT. Here we analyze the
quantum states generated in each of these stages and
evaluate their entanglement using the Groverian measure.

\subsection{Entanglement Generated by the QFT Procedure}

Here we evaluate the time evolution of the Groverian entanglement
during the QFT process, 
shown in Fig. 
\ref{fig:1}.
The Groverian measure is evaluated 
after each operation of the 
$B_{k,m}$ operator.
The $A_k$ operators are local and
do not change the entanglement,
We first perform this analysis for general quantum states and then
focus on the specific quantum states that appear in the factoring
algorithm.

\subsubsection{QFT Applied on General Quantum States}

To examine the effect of QFT on the Groverian entanglement we
construct an ensemble of random product states as well as random
states of $n$ qubits.
The state of each qubit in the
random product states is described by 
Eq. 
(\ref{eq:e_j})
where 
$0 \le \theta_k < \pi$
and
$0 \le \gamma_k < 2 \pi$
are chosen randomly.
The random states are drawn from an isotropic distribution
in the $2^L$-dimensional Hilbert space
\cite{Shimoni2004}.
These states turn out to be highly entangled.

\begin{figure}
\includegraphics[width=8.5cm]{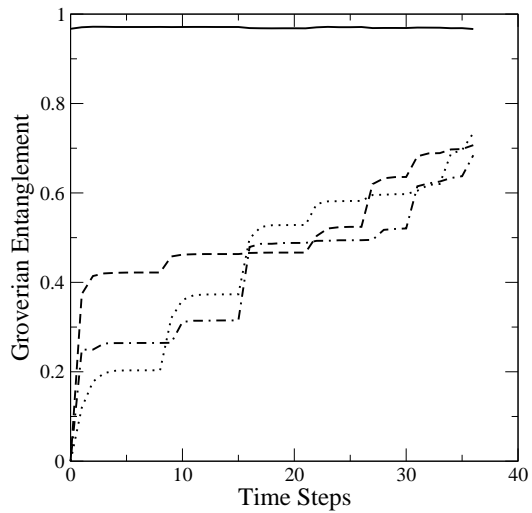}
\caption{The Groverian measure of entanglement for states created
during the operation of the QFT on three randomly chosen tensor product states
(dashed, dotted and dashed-dotted) 
as well as on a single random state (solid line). 
All the states are 
of nine qubits.}

\label{fig:2}
\end{figure}

In Fig. 
\ref{fig:2} 
we present the time evolution of the 
Groverian measure during the processing of QFT on
three random product states as well as on a random state
of nine qubits.
For the random product states one observes that during most
time steps the entanglement remains unchanged.
Most of the variation takes place at specific times,
common to all the different states.
Clearly, the entanglement is generated by the
controlled phase operators
$B_{k,m}$.
The large variations 
in $G(\psi)$
are found to take place when
$|m-k|$ is small,
namely when $B_{k,m}$
is applied on pairs of adjacent qubits.
The Groverian measure during the operation of QFT on 
a highly entangled random state is also shown in
Fig. 
\ref{fig:2}.
It exhibits only small variations with no obvious regularity.

\subsubsection{QFT Within Shor's Factoring Algorithm}

\begin{figure}
\includegraphics[width=8.5cm]{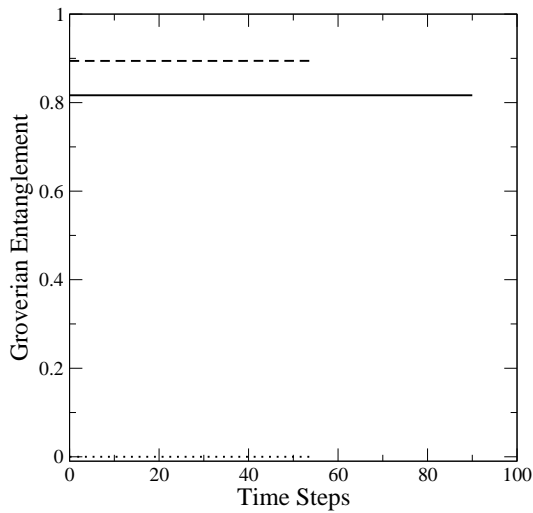}
\caption{The Groverian measure of entanglement for states created 
during the QFT stage of Shor's factoring algorithm.
The solid line shows the factorization of $N=91$ using $y=41$.
The dotted line (with zero entanglement)
shows the factorization of $N=33$ using $y=23$.
The dashed line shows the factorization of $N=33$ using $y=4$.}
\label{fig:3}
\end{figure}

In
Fig. 
\ref{fig:3} 
we present the time evolution of the Groverian measure
during QFT, when it is applied on states
obtained from the pre-processing stage of Shor's factoring 
algorithm.
The different lines correspond to the factorization process of 
different numbers. 
Surprisingly,
for all numbers that we have tested,
the entanglement was essentially unchanged throughout
the process,
as implied by the horizontal lines.
This is
in contrast to the behavior observed when QFT is applied
to general quantum states.

A special property of the states generated by the pre-processing is
that they are periodic. This motivated us to 
examine the time evolution of the Groverian measure 
during QFT of general periodic states.
The state
$\sum_m |l+mr\rangle$ 
(up to normalization factor) 
is a periodic state of $L$ qubits, 
with period $r$ and shift $l$. 
The summation is over all integers $m$ such that
$0 \leq l+mr \leq q-1$, 
where $q=2^L$. 
It was found that the Groverian measure essentially does not change
during the QFT process of such states, 
and that the changes which do occur vanish exponentially 
with the number of qubits.
The value of the Groverian measure for these states 
depends almost solely on the odd part of the period $r$.
More precisely,
for a periodic state with period 
$r=2^M d$ (where $d$ is odd), 
we obtain 
$P_{\max} \simeq 1/d$. 
This is easy to explain for states with a period $r=2^M$,
which are known to be tensor product states. 
For these states $d=1$, thus the correct result of $P_{\max}=1$
is obtained.
For general periodic states 
we do not have an analytical derivation of the expression for
$P_{\max}$. 

\subsection{Entanglement in the Pre-processing Stage}

\begin{figure}
\includegraphics[width=8.5cm]{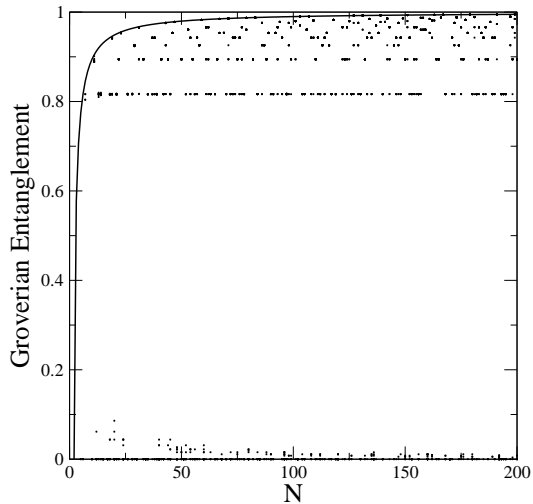}
\caption{The Groverian measure of entanglement for the 
states created by the
pre-processing stage of Shor's algorithm. 
Each dot corresponds to a single choice of $2<N\leq200$ and $1<y<N-1$.}
\label{fig:4}
\end{figure}

Having found that the QFT stage of Shor's algorithm does not alter the 
entanglement of states created by the pre-processing
stage, it is clear that all the entanglement is produced during
pre-processing.
We have evaluated this entanglement generated during the
factoring process of all the integers in the range
$3 \le N \le 200$.
To factorize an integer,
$N$, one has to choose another integer $1 <y < N-1$.
In our analysis, we examined all possible choices within
this range, and for each of them we applied
the pre-processing stage 
as described in Sec.
\ref{sec:algorithm}.
At the end of the pre-processing stage we 
evaluated the Groverian measure of the resulting state of the main register,
following a measurement of the auxiliary register.
In Fig. 
\ref{fig:4}
we present the Groverian measure for the states obtained after
pre-processing vs. $N$ for 
$3 \le N \le 200$.
Each dot represents the 
Groverian measure after pre-processing for the integer $N$
and for a specific choice of $1 <y < N-1$.
The solid line represents the function
$\sqrt{1-1/(2N)}$.
We observe that all the dots are below this line, 
which resembles the upper bound of the 
Groverian measure, namely that for any state
$|\psi\rangle$ of $L$ qubits
$G(\psi) \leq \sqrt{1-1/2^L}$.

Additionally, there are many  
values of $N$ and choices of $y$ 
for which the Groverian measure is $G=0$,
namely
the factoring process does not involve any entanglement.
For these particular choices, it should thus 
be possible to perform the factoring of $N$ efficiently
using a classical algorithm
\cite{Aharonov1996}.
We find that for some of the pairs of $N$ and $y$
which produce no entanglement,
GCD$(N,y) \ne 1$,
thus a divisor of $N$ can be easily
found classically.
The rest of these pairs are found to satisfy
$y^{2^n} = 1 \ {\rm mod}\ N$, 
for some integer $n$,
which means that 
GCD$(y^{2^{n-1}} + 1 ,N)$ 
or 
GCD$(y^{2^{n-1}} - 1 ,N)$
are divisors of $N$,
which can be easily found by classical algorithms.
We thus find that in cases in which 
no entanglement is produced by the 
quantum algorithm,
it offers no speedup compared to classical algorithms.
This is consistent with the assumption that the entanglement
generated by a quantum algorithm is correlated with the
speedup it provides.

\section{Discussion}
\label{sec:discussion}


It is found that the states prepared by
the pre-processing stage of Shor's algorithm,
like all periodic states,
exhibit the property
that their Groverian entanglement does 
not change throughout the QFT stage.
One may take the view
that the Groverian entanglement somehow represents
the amount of quantum information present in a quantum state.
This is rather like the von Neumann entropy. 
Taking this view, our result may seem
natural
because the information needed to perform the factoring
is already present after the pre-processing stage.
The QFT only rearranges the information such that it can 
be extracted by measurement.

It is found that
the Groverian measure of the states
generated by Shor's algorithm is lower than
that of random states, which are
almost maximally entangled, with 
$G(\psi) \simeq \sqrt{1- 1/q}$
\cite{Biham2003,Shimoni2004}. 
Yet, the maximal entanglement created by the 
algorithm exhibits the same functional behaviour,
where $q$ is raplaced by $2N$.

Considering the fact that Shor's algorithm is exponentially 
faster than its known classical counterparts, 
it is expected to use all the entanglement available.
Thus, our result provides further indication 
that classical algorithms are unlikely to
perform factoring in polynomial time.

Unlike Shor's algorithm, 
Grover's search algorithm
is only polynomialy more efficient than its classical counterparts
\cite{Grover1996,Grover1997}.
Grover's algorithm also
creates entanglement,
which is bound by a constant lower than unity
\cite{Biham2003}.

A different approach to the analysis of the entanglement generated
by Shor's factoring algorithm was presented in Ref.
\cite{Kendon2005}, where
the bi-partite entaglement between the  
main register and the auxiliary
register was evaluated
during both the pre-processing and QFT stages, using the negativity
\cite{Peres1996,Karol1998}
as an entanglement measure.
It was found that the entanglement is primarily generated during
the pre-processing stage, in agreement with our results.

\section{Summary}
\label{sec:summary}

The quantum states created during the operation 
of Shor's factoring algorithm have been analyzed and
the entanglement in these states was evaluated using the
Groverian measure.
It was found that the entanglement is generated during the
pre-processing stage and remains unchanged during
the QFT stage.
It was shown that
the latter
feature is unique to periodic states, 
such as those obtained from the pre-processing stage,
while
QFT does affect the entanglement
of general quantum states. 
Another interesting feature is that
the entanglement is found to be correlated with the speedup 
achieved by the quantum algorithm compared to classical 
algorithms.
This means that the cases where no entanglement is created
are those in which classical factoring is efficient. 



\begin{thebibliography}{10}

\bibitem{Shor1994}
{P.W. Shor},  in {\em {Proceedings of the 35th Annual Symposium on the
  Foundations of Computer Science}}, edited by {S. Goldwasser} ({IEEE Computer
  Society}, {Los Alamitos, CA}, 1994), p.\ 124.

\bibitem{Grover1996}
{L. Grover},  in {\em {Proceedings of the Twenty-Eighth Annual Symposium on the
  Theory of Computing }} ({ACM Press}, {New York}, {1996}), p.\ 212.

\bibitem{Grover1997}
{L. Grover}, Phys. Rev. Lett. {\bf 79},  325  (1997).

\bibitem{Deutsch1992}
{D. Deutsch and R. Jozsa}, Proc. R. Soc. London, A {\bf 439},  553  (1992).

\bibitem{Jozsa1997}
{R. Josza}, {Quantum Algorithms and the Fourier Transform}, {e-print
  quant-ph/9707033}.

\bibitem{Jozsa2003}
{R. Jozsa and N. Linden}, Proc. R. Soc. London, A {\bf 459},  2011  (2003).

\bibitem{Vidal2003}
{G. Vidal}, Phys. Rev. Lett. {\bf 91},  147902  (2003).

\bibitem{Aharonov1996}
{D. Aharonov and M. Ben-Or},  in {\em {Proceedings of the 37th Annual Symposium
  on the Foundations of Computer Science}}, edited by {S. Goldwasser} ({IEEE
  Computer Society}, {Los Alamitos, CA}, 1996), p.\ 46.

\bibitem{Biham2002}
{O. Biham, M.A. Nielsen and T. Osborne}, Phys. Rev. A {\bf 65},  062312
  (2002).

\bibitem{Shimoni2004}
{Y. Shimoni, D. Shapira and O. Biham}, Phys. Rev. A {\bf 69},  062303  (2004).

\bibitem{Vedral1997}
{V. Vedral, M.B. Plenio, M.A. Rippin and P.L. Knight}, Phys. Rev. Lett. {\bf
  78},  2275  (1997).

\bibitem{Vedral1998}
{V. Vedral and M.B. Plenio}, Phys. Rev. A {\bf 57},  1619  (1998).

\bibitem{Vedral1997a}
{V. Vedral, M. B. Plenio, K. Jacobs, and P. L. Knight}, Phys. Rev. A {\bf 56},
  4452  (1997).

\bibitem{Miyake2001}
{A. Miyake and M. Wadati}, Phys. Rev. A {\bf 64},  042317  (2001).

\bibitem{Biham2003}
{O. Biham, D. Shapira and Y. Shimoni}, Phys. Rev. A {\bf 68},  {022326}
  (2003).

\bibitem{Kendon2005}
{V.M. Kendon and W.J. Munro}, {Entanglement and its role in Shor's algorithm},
  {e-print quant-ph/0412140}.

\bibitem{Peres1996}
{A. Peres}, Phys. Rev. Lett. {\bf 77},  1413  (1996).

\bibitem{Karol1998}
{K. $\dot{\rm Z}$yczkowski, P. Horodecki, A. Sanpera, M. Lewenstein}, Phys.
  Rev. A {\bf 58},  883  (1998).

\end{thebibliography}
\end{document}